\documentclass[epj]{svjour}

\usepackage{graphicx}
\usepackage{fancyhdr}

\def\mytitle{My title} 
\def\myauthors{My name}  
\def\mytype{My type of session}
\def\mysession{My session}
\def\mytitle{Effective superpotential and partial breaking of $\mathcal{N}=2$ supersymmetry} %Put your title here!
\def\myauthors{Kazunobu Maruyoshi}    %Put your name here!
\def\mytype{Contributed Talk}    
\def\mysession{Theoretical Models}

\begin{document}
\title{Effective superpotential and partial breaking of $\mathcal{N}=2$ supersymmetry}
\author{Kazunobu Maruyoshi
%\inst{1}
\thanks{\emph{Email:} maruchan@sci.osaka-cu.ac.jp}%
% \and
% Second author\inst{2}% etc
}                     % Do not remove
\institute{Department of Mathematics and Physics, Graduate School of Science, 
Osaka City University, Osaka, 558-8585, Japan
%\and the second institute
%address here
}
\date{}
\abstract{
We consider the effective superpotential of $\mathcal{N}=2$, $U(N)$ gauge model 
where $\mathcal{N}=2$ supersymmetry is spontaneously broken to $\mathcal{N}=1$.
By the computation of loop diagrams, we obtain a formula for the effective superpotential
which is deformed from the well-known form of the effective superpotential of $\mathcal{N}=1$, $U(N)$ gauge model with 
a tree level superpotential.
 % end of PACS codes
} %end of abstract
\maketitle
%%%%%%%%%%%%%%%%%%%%%%%%%%%%%%%%%%%%%%%%%%%%%%%%%%%%%%%%%%%%%%%%%%%%%%%%%%%%%%%%%%%%%%%%%%%%%%%%%%%%%%%%%%%%%%%%
\section{Introduction}
\label{intro}
  One of the important progresses in recent years in theoretical physics is the discovery of supersymmetry.
  Various investigations have been made 
  on $\mathcal{N}=1$ and $\mathcal{N}=2$ supersymmetric Yang-Mills theory in four dimensions.
  In particular, based on the works \cite{Vafa,CIV,DV1}, it has been suggested \cite{DV2}
  that the computation of the effective superpotential of $\mathcal{N}=2$ supersymmetric $U(N)$ gauge theory
  \textit{softly} broken to $\mathcal{N}=1$ by adding a superpotential 
  reduces to that of the free energy of the bosonic one matrix model.
  
  On the other hand, in view of the fact 
  that superstring theories produce, in some backgrounds, extended supersymmetry in four dimensions 
  and have no adjustable parameter, 
  it is natural to consider spontaneous breaking of the extended supersymmetry
  so as to obtain more realistic $\mathcal{N}=1$ supersymmetric models.
  The possibility of the partial breaking has been first pointed out in \cite{HP}
  by the argument based on the supercurrent algebra 
  which has been modified by an additional space-time independent term:
    \begin{eqnarray}
    \{ \bar{Q}^J_{\dot{\alpha}}, \mathcal{S}_{\alpha I}^\mu \}
     =     2 (\sigma^\nu)_{\alpha \dot{\alpha}} T_\nu^\mu(x) \delta_I^J + (\sigma^\mu)_{\alpha \dot{\alpha}} C_I^J, 
           \nonumber
    \end{eqnarray}
  where $\mathcal{S}_{\alpha I}^\mu$ and $T_\nu^\mu$ are, respectively, the supercurrents and the energy momentum tensor.
  In \cite{APT} and \cite{FIS1,FIS2,FIS3,IMS}, $\mathcal{N}=2$, $U(1)$ and $U(N)$ gauge models have been constructed, 
  establishing this modification of the algebra by introducing the electric and magnetic Fayet-Iliopoulos term.
  (See also \cite{sugra} for supergravity.)
  
  So, it is natural to ask the following questions: 
  can the effective superpotential of the model 
  with partially as well as spontaneously broken $\mathcal{N}=2$ supersymmetry
  be deformed from that of \textit{softly} broken case \cite{DV2} ?
  Also, how is the relation to the matrix model modified by spontaneous breaking ?
  In \cite{IM2,IM3}, we make a first analysis on the effective superpotential of the model \cite{FIS1,FIS2}.
  In this paper, we briefly review this analysis.

  In section \ref{sec:model}, we introduce the model \cite{FIS1,FIS2} and collect some properties of this model.
  Then, we will consider purely gauge theoretic approach to (the matter induced part of) the effective superpotential 
  which is obtained by integrating out the massive adjoint superfield $\Phi$:
    \begin{eqnarray}
    e^{i \int d^4 x (d^2 \theta W_{eff} + h.c. + ({\rm D-term}))}
    & &    ~~~~~~~~~~~~~
           \nonumber \\
    =    \int \mathcal{D} \Phi \mathcal{D} \bar{\Phi} e^{i \int d^4 \mathcal{L}_{\mathcal{N}=1}^{\mathcal{F}}}.
    & &
           \label{Weff}
    \end{eqnarray}  
  where $\mathcal{L}_{\mathcal{N}=1}^{\mathcal{F}}$ is given in the section \ref{sec:model}.
  Our approach is the explicit computation of loop diagrams based on \cite{DGLVZ}.
  We will see this in section \ref{sec:diagram}.

%%%%%%%%%%%%%%%%%%%%%%%%%%%%%%%%%%%%%%%%%%%%%%%%%%%%%%%%%%%%%%%%%%%%%%%%%%%%%%%%%%%%%%%%%%%%%%%%%%%%%%%%%%%%%%%%%%
\section{The model}
\label{sec:model}
  The lagrangian of the model \cite{FIS1,FIS2} is, in $\mathcal{N}=1$ superspace formalism,
    \begin{eqnarray}
    \mathcal{L}_{\mathcal{N}=2}^{\mathcal{F}}   
    &=&    \int d^4 \theta 
           \left[
         - \frac{i}{2} {\rm Tr} 
           \left(  \bar{\Phi} e^{ad V} 
           \frac{\partial \mathcal{F}(\Phi)}{\partial \Phi}
         - h.c.
           \right)
         + \xi V^0 
           \right] 
           \nonumber \\
    & &  + \Bigg[
           \int d^2 \theta
           \Bigg(
         - \frac{i}{4} 
           \frac{\partial^2 \mathcal{F}(\Phi)}{\partial \Phi^a \partial \Phi^b}
           \mathcal{W}^{\alpha a} \mathcal{W}^b_{\alpha}
           \nonumber \\
    & &    ~~~~~~~~~~~~~~
         + e \Phi^0
         + m \frac{\partial \mathcal{F}(\Phi)}{\partial \Phi^0}
           \Bigg)
         + h.c.
           \Bigg],      
           \nonumber
    \end{eqnarray}
  where $V$ and $\Phi$ are vector and chiral superfields
  whose fermionic components are $\lambda$ and $\psi$ respectively.
  These superfield $\Psi =\{ V, \Phi \}$ is written as $\Psi = \sum_{a=0}^{N^2 - 1} \Psi^a t_a$
  where $t_a$ are the generators ($a = 0$ refers to the overall $U(1)$ generator).
  Theoretical inputs are the electric and magnetic Fayet-Iliopoulos terms 
  which are two vectors or a rank two symmetric tensor in the isospin space 
  and are parameterized by the three real parameters $e, m, \xi$ in the $\mathcal{N}=1$ superspace formalism we employ.
  In addition, the model contains an arbitrary input function $\mathcal{F}(\Phi)$, 
  which we refer to as a prepotential.
  Its prototypical form is a single trace function of a polynomial in $\Phi$:
    \begin{eqnarray}
    \mathcal{F}(\Phi)
     =     \sum_{k=1}^{n+1} \frac{g_k}{(k+1)!} {\rm Tr} \Phi^{k+1},
           ~~
           {\rm deg} \mathcal{F} = n+2.
           \nonumber
    \end{eqnarray}
    
  While this action is shown to be invariant under the $\mathcal{N}=2$ supersymmetry
  transformations \cite{FIS1,FIS2},  the vacuum breaks half of 
  the $\mathcal{N}=2$ supersymmetries.
  Extremizing the scalar potential, we obtain the following condition,  
    \begin{eqnarray}
    \langle \frac{\partial^2 \mathcal{F}(\Phi)}{\partial \Phi^0 \partial \Phi^0} \rangle 
     =   - \left( \frac{e}{m} \pm \frac{i \xi}{m} \right)
           \nonumber
    \end{eqnarray}
  which is a polynomial of order $n$ and this determines the expectation value of the scalar field.
  In these vacua, the combination of the fermions, $\lambda \mp \psi$, becomes massive, 
  on the other hand, $\lambda \pm \psi$ is massless, 
  whose overall $U(1)$ component is Nambu-Goldstone fermion \cite{FIS1,FIS2}.
  In order to obtain the action on the vacua, we, therefore, have to redefine the superfields $V$ and $\Phi$ 
  such that the fermionic components of them mix as above.
  In \cite{Fujiwara}, the action on the vacua has been obtained by taking account 
  of this point and that Fayet-Iliopoulos D-term can be included in the superpotential,
    \begin{eqnarray}
    \mathcal{L}_{\mathcal{N}=1}^{\mathcal{F}}   
    &=&    \int d^4 \theta 
           \left[
         - \frac{i}{2} {\rm Tr} 
           \left(  \bar{\Phi} e^{ad V} 
           \frac{\partial \mathcal{F}(\Phi)}{\partial \Phi}
         - h.c.
           \right) 
           \right] 
           \nonumber \\
    & &  + \Bigg[
           \int d^2 \theta
           \left(
         - \frac{i}{4} 
           \frac{\partial^2 \mathcal{F}(\Phi)}{\partial \Phi^a \partial \Phi^b}
           \mathcal{W}^{\alpha a} \mathcal{W}^b_{\alpha}
         + W(\Phi)
           \right)
           \nonumber\\
    & &    ~~~~~~~~~~~~~~~~~
         + h.c.
           \Bigg],      
           \nonumber
    \end{eqnarray}
  where 
    \begin{eqnarray}
    W(\Phi)
     =     {\rm Tr} \left[
           2 (e \pm i \xi) \Phi + m \sum_{k=1}^{n+1} \frac{g_k}{k!} \Phi^k
           \right]
           \label{Wtree}
    \end{eqnarray}
  which is the single trace function of degree $n+1$.
  In (\ref{Wtree}), we have redefined $e, m, \xi$ such that they include the factor $1/\sqrt{2 N}$ 
  which comes from the overall $U(1)$ generator $t_0 = 1_{N \times N}/\sqrt{2 N}$.
  It is, also, understood that $V$ and $\Phi$ has been redefined as noted above.
  
  Before going on to the analysis of the effective action, 
  let us mention the relation with $\mathcal{N}=1$ super Yang-Mills theory with a tree level superpotential
  which has been considered by \cite{Vafa,CIV,DV1,DV2}.
  The lagrangian $\mathcal{L}_{\mathcal{N}=1}^{\mathcal{F}}$ is to be compared with
  that of the $\mathcal{N}=1$, $U(N)$ gauge model with a single trace tree level superpotential $W(\Phi)$:
    \begin{eqnarray} 
    \mathcal{L}_{\mathcal{N}=1}
    &=&    \int d^4 \theta 
           {\rm Tr} \bar{\Phi} e^{ad V} \Phi
           \nonumber \\
    & &  + \left[
           \int d^2 \theta
           {\rm Tr} \left(
           i \tau \mathcal{W}^\alpha \mathcal{W}_\alpha
         + W(\Phi)
           \right)
         + h.c. 
           \right].
           \label{S1}
    \end{eqnarray}
  $\tau$ is a complex gauge coupling $\tau = \theta/2 \pi + 4 \pi i / g^2$.
  In \cite{FIS1}, it is checked that the second supersymmetry reduces 
  to the fermionic shift symmetry \cite{CDSW} by taking the Fayet-Iliopoulos parameters to be infinity.
  Also, the lagrangian $\mathcal{L}_{\mathcal{N}=1}^{\mathcal{F}}$ in fact reduces 
  to $\mathcal{L}_{\mathcal{N}=1}$ explicitly
  in the limit $m, e, \xi \rightarrow \infty$ with $\tilde{g}_\ell \equiv m g_\ell$ ($\ell \geq 2$) fixed \cite{Fujiwara}.
  We refer to this limit as $\mathcal{N}=1$ limit.
  
  If we keep $e,m,\xi$ finite, the behavior of the quantum effective lagrangian may differ
  from that of \cite{Vafa,CIV,DV1,DV2}.
  Understanding of this behavior is the central issue in our investigation.    
                      
%%%%%%%%%%%%%%%%%%%%%%%%%%%%%%%%%%%%%%%%%%%%%%%%%%%%%%%%%%%%%%%%%%%%%%%%%%%%%%%%%%%%%%%%%%%%%%%%%%%%%%%%%%%%%%%%%5
\section{Diagrammatic analysis of the effective superpotential}
\label{sec:diagram} 
  In this subsection, we review the diagrammatical computation of the effective superpotential \cite{IM2}.
  
  Let us take $\mathcal{W}^\alpha$ (or $V$) as the background field.
  The simplest background is that consisting of 
  a vanishing gauge field $A_\mu$ and a constant gaugino $\lambda^\alpha$, 
  which satisfies $\{ \lambda^\alpha, \lambda^\beta \} = 0$ \cite{AFH,IM2}.
  This configuration implies that traces of more than two $\mathcal{W}^\alpha$ vanish.
  Furthermore, for simplicity,  we consider the case of unbroken $U(N)$ gauge group
  and choose $\left< \Phi \right> = 0$ by setting the coupling as $g_1 = - (e \pm i \xi)/m$.
  Therefore, the result of the diagrammatical computation can be written 
  in terms of the coupling constants $\tilde{g}_\ell$, the Fayet-Iliopoulos parameter $m$, 
  the glueball superfield $S \equiv - {\rm Tr} \mathcal{W}^\alpha \mathcal{W}_\alpha/64\pi^2 $ 
  and the overall $U(1)$ field strength $w^\alpha \equiv {\rm Tr} \mathcal{W}^\alpha /8 \pi$.
  
  We start from (\ref{Weff}) and integrate $\bar{\Phi}$.
  This is easily done by setting the anti-holomorphic couplings $\bar{g}_k=0$ for $k\geq3$.
  (The effective superpotential is holomorphic and does not include the anti-holomorphic couplings.
  Hence, this choice does not change the result.)
  In this choice, if we focus on the $\bar{\Phi}$-dependent terms in $\mathcal{L}_{\mathcal{N}=1}^{\mathcal{F}}$,
  we can see that the $\bar{\Phi}$-integral becomes Gaussian integral and we get
    \begin{eqnarray}
    \frac{1}{16 \bar{g}_2}
    \left(
    \bar{g}_1 \Phi - \frac{\partial \mathcal{F}}{\partial \Phi}
    \right)
    \left(
    - \frac{2 m}{\nabla^2} + \frac{i \Phi}{4} 
    \right)^{-1}
    \left(
    \bar{g}_1 \Phi - \frac{\partial \mathcal{F}}{\partial \Phi}
    \right)
    & &    \nonumber \\
     =     \frac{({\rm Im} g_1)^2}{8 \bar{\tilde{g}}_2} \Phi \nabla^2 \Phi
         + V(\Phi),~~~
    & &    
           \label{SPhi2}
    \end{eqnarray}
  where $V(\Phi)$ denotes the higher order interaction terms.
  Note that $V(\Phi)$ is $\mathcal{O}(1/m)$.
  Also, in $\mathcal{L}_{\mathcal{N}=1}^{\mathcal{F}}$, there are $\bar{\Phi}$-independent terms
    \begin{eqnarray}
    & &    \int d^2 \theta {\rm Tr}
           \sum_{k=2}^{n+1}
           \left(
           \frac{\tilde{g}_k}{k!} \Phi^k
         - \sum_{s=0}^{k-1} \frac{i g_k}{4 k!} 
           (\mathcal{W}^\alpha \Phi^s \mathcal{W}_\alpha \Phi^{k-1-s})
           \right).
           \nonumber \\
    & &    \label{SPhi}
    \end{eqnarray}
  
  From (\ref{SPhi2}) and (\ref{SPhi}), we read off the Feynman rule.
  In eq. (\ref{SPhi}), there is a linear term in $\Phi$.
  However, it is understood that the generating functional has a renormalized perturbation expansion
  in which a nonvanishing tadpole is always canceled by a nonvanishing value of the source coupled to $\Phi$.
  This implies that the linear term can in practice be ignored.
  Collecting the quadratic terms, the propagator in momentum space can be written as
    \begin{eqnarray}
    \Delta(p, \pi)
     =     \int_0^{\infty} d s 
           e^{-s (p^2 + m' + \frac{1}{2} ad\mathcal{W}^\alpha \pi_\alpha - i g'_3 M}),
           \nonumber
    \end{eqnarray}
  where we have defined $ m'= a^2 \tilde{g}_2$, $g'_3 = a^2 g_3/12$ and $a^2=\bar{\tilde{g}}_2/({\rm Im} g_1)^2$.
  The Grassmann momentum $\pi^\alpha $ is Fourier transformation of
  superspace coordinate $\theta^\alpha$ and  the matrix $M$ is
    \begin{eqnarray}
    M_{abcd}
     =     (\mathcal{W}^\alpha \mathcal{W}_\alpha)_{da} \delta_{bc} 
         + (\mathcal{W}^\alpha \mathcal{W}_\alpha)_{bc} \delta_{da} 
         + \mathcal{W}_{da}^\alpha \mathcal{W}_{\alpha bc}.
           \label{M}
    \end{eqnarray}
  This matrix is not present in the propagator of \cite{DGLVZ}.
  
  The interaction terms in (\ref{SPhi2}) and (\ref{SPhi}) are divided into the following three types:
    \begin{eqnarray}
    ~~{\rm type~I}
    & &    ~~~~~~~
           \frac{\tilde{g}_k a^k}{k!} {\rm Tr} \Phi^k, 
           ~~~~~~~~~~
           k = 3, \ldots, n+1.
           \nonumber \\
    ~~{\rm type~I\hspace{-.1em}I}
    & &    ~~~~
         - \sum_{s=0}^{k-1}
           \frac{i g_k a^{k-1}} {4 k!} {\rm Tr} (\mathcal{W}^\alpha \Phi^s \mathcal{W}_\alpha \Phi^{k-1-s}),
           \nonumber \\
    & &    ~~~~~~~~~~~~~~~~~~~~~~~~~~~~~~~~
           k = 4, \ldots, n+1.
           \nonumber \\
    {\rm type~I\hspace{-.1em}I\hspace{-.1em}I}
    & &    ~~~~~~~
           {\rm each~ term~ in~}V(\Phi)
           \nonumber
    \end{eqnarray}
  Type I vertices come from the first term in (\ref{SPhi}) and are, therefore, also present in \cite{DGLVZ}.
  Type I\hspace{-.1em}I vertices which come from the second term in (\ref{SPhi}) 
  and type I\hspace{-.1em}I\hspace{-.1em}I vertices are not present in \cite{DGLVZ}.
  We can see below that these new ingredients, $M$ term in the propagator 
  and type I\hspace{-.1em}I and type I\hspace{-.1em}I\hspace{-.1em}I vertices
  do contribute to the effective superpotential.
  
  The perturbative part of the effective superpotential is derived by computing the amplitude of all the diagrams.
  Let us first see that the non-planar diagrams do not contribute.
  For a given diagram,  we denote by $V$ the number of vertices,
  by $P$ the number of propagators and by $h$ the number of index loops.
  There are $V$ sets of chiral superspace integrations from $V$ vertices.
  One of them becomes the chiral superspace integration over the effective superpotential,  
  and the number of remaining $\pi^\alpha$ momentum integrations is $P - V + 1$.
  These Grassmann integrations must be saturated by $\frac{1}{2} ad \mathcal{W}^\alpha \pi_\alpha$ terms 
  in the propagators.
  Furthermore, we can freely insert $\mathcal{W}^\alpha$ 
  both from the $M$ terms in the propagators and from the type I\hspace{-.1em}I vertices.
  If we denote the number of these additional insertions by $2 \alpha$, 
  the total number of $\mathcal{W}^\alpha$ insertions is $2 (P - V + 1 + \alpha)$.
  On the other hand, one index loop can accommodate at most two $\mathcal{W}^\alpha$.
  Thus we have $h \geq P - V + 1 + \alpha$.  
  This implies that only the planar diagrams contribute to the effective
  superpotential as the Euler number of the diagram is $\chi = V - P + h$.
  
  Let us consider the $L$-loop planar diagrams with $P$ propagators in which all vertices are type I
  (as we have seen above, $L = P -V + 1 = h -1$).
  For a moment, we ignore the $M$ term of (\ref{M}).
  The calculation is then the same as that of \cite{DGLVZ}: 
  as we have seen in the last paragraph, we have exactly $2L$ $\mathcal{W}^\alpha$ insertions.
  There are two possibilities for these $\mathcal{W}^\alpha$ insertions.
  The one is to keep one of the index loops empty, filling the remaining index loops with two $\mathcal{W}^\alpha$.
  This yields $N S^L$ term.
  The other is to fill each of two index loops chosen with single $\mathcal{W}^\alpha$, 
  which yields $S^{L-1} w^\alpha w_\alpha$ terms.
  After calculating $p$ momentum and $\pi^\alpha$ momentum integrals, we perform the Schwinger parameter integrals.
  The most important fact is that the Schwinger parameter integrals simply reduce to \cite{DGLVZ,IM2}
    \begin{eqnarray}
    \left(
    \prod_{i=1}^{P} \int d s_i 
    \right)
    e^{-(\sum s_i)m'}
     =     \frac{1}{m'^P}
    \end{eqnarray}
  Clearly, this procedure is universal to every $L$-loop planar diagram with $P$ propagators 
  and the amplitude is given by
    \begin{eqnarray}
    \frac{1}{4^L m'^P} \{ N (L+1) S^L
    + L(L + 1) S^{L-1} w^\alpha w_\alpha \},
    \label{amp1}
    \end{eqnarray}
  up to the multiplications by the symmetric factor and by the coupling constants.
  The factor $L+1$ of the first term comes from the choice of the empty index loop, 
  and $L(L + 1)$ of the second term is the combination of inserting two $\mathcal{W}^\alpha$ 
  into different index loops.
  
  As we have seen above, (\ref{amp1}) is universal to every $L$-loop planar diagram with $P$ propagator, 
  but each diagram has different symmetric factor and coupling constants.
  Adding the symmetric factors and the coupling constants which come from all the possible $L$-loop planar diagrams,
  we obtain the $L$-loop contribution to the effective superpotential:
    \begin{eqnarray}
    W_{eff}^{(L)}
     =     N \frac{\partial F^{(L)}}{\partial S}
         + \frac{\partial^2 F^{(L)}}{\partial S^2} w^\alpha w_\alpha,
           \label{Weffh1}
    \end{eqnarray}
  where $F^{(L)}$ include non-universal factor, 
  the symmetric factors and the coupling constants which are from various planar $L$-loop diagrams
  and universal factor, $S^{L+1}$.
  As in \cite{DGLVZ,IM2}, $F^{(L)}$ agrees with the $L$-loop contribution 
  to the planar free energy of the bosonic one matrix model.
  This explains that, in the theory $\mathcal{L}_{\mathcal{N}=1}$, 
  the calculation of the effective superpotential of the gauge theory reduces to that of the matrix model \cite{DV2}.
  In our model, as we will see below, 
  we can obtain the additional contributions
  by taking into account of the $M$ term in the propagator 
  and the type I\hspace{-.1em}I and the type I\hspace{-.1em}I\hspace{-.1em}I vertices.
  Let us see this in below.
  
  There are three types of corrections to (\ref{Weffh1}).
  The one is due to the presence of the $M$ terms in the propagators.
  The others are due to the type I\hspace{-.1em}I vertices and type I\hspace{-.1em}I\hspace{-.1em}I vertices, 
  which is, respectively, obtained by replacing one of the type I vertices 
  in (\ref{Weffh1}) by the corresponding type I\hspace{-.1em}I vertex and type I\hspace{-.1em}I\hspace{-.1em}I vertex
  and by summing over all possibilities.
  We consider them in order.
  
  First of all, let us see the effects of the $M$ term, namely, (\ref{M}).
  It plays a role of inserting two $\mathcal{W}^\alpha$ further.
  Thus we will obtain terms which are proportional to $S^{L+1}$.
  Note that we cannot insert more than two $\mathcal{W}^\alpha$  because, in such case, 
  at least one of the index loops has more than two insertions of $\mathcal{W}^\alpha$.
  For the parts contributing to $N S^L$ in (\ref{amp1}), which have an empty index loop, 
  we can further insert $\mathcal{W}^\alpha \mathcal{W}_\alpha$ from the first two terms in (\ref{M}).
  The parts contributing to the second term of (\ref{amp1}) can receive 
  further insertions of $\mathcal{W}^\alpha$ as well.
  They have two index loops with a single $\mathcal{W}^\alpha$ insertion, for which we can exploit the last term of $M$.
  Considering these $M$ term insertions, we obtain the first modification to (\ref{Weffh1}) \cite{IM2}
    \begin{eqnarray}
    - \frac{16 \pi^2 i P \tilde{g}_3 S}{(L + 1) m \tilde{g}_2} 
           \left(
           \frac{\partial F^{(L)}}{\partial S}
           \right).
    \label{Wefftype1}
    \end{eqnarray}
  
  Secondly, let us consider that one of the type I vertices ${\rm Tr} \Phi^\ell$ of a particular $L$-loop planar diagram
  is replaced by the type I\hspace{-.1em}I vertex.
  The $\ell$-th order type I\hspace{-.1em}I vertex in $\Phi$ is
    \begin{eqnarray}
    {\rm Tr}
    (2 \mathcal{W}^\alpha \mathcal{W}_\alpha \Phi^\ell 
    + \mathcal{W}^\alpha \Phi \mathcal{W}_\alpha \Phi^{\ell - 1} 
    + \dots 
    + \mathcal{W}^\alpha \Phi^{\ell - 1} \mathcal{W}_\alpha \Phi),
    \nonumber
    \end{eqnarray}
  We have omitted the overall factors and the trace.
  The first term inserts two $\mathcal{W}^\alpha$ into an index loop while 
  the remainder insert them into two different index loops.
  As above, if we consider only the type I vertices, we obtain $2L$ $\mathcal{W}^\alpha$ insertions.
  We can therefore use the type I\hspace{-.1em}I only once in a diagram.
  When this is done, insertion of the $M$ term from the propagator is disallowed.
  In any $L$-loop diagram, the effect of changing a vertex from type I to type I\hspace{-.1em}I 
  is equivalent to considering a corresponding $L$-loop contribution in the first term in (\ref{Weffh1}) 
  and changing the coupling constant by \cite{IM2}
    \begin{eqnarray}
    \tilde{g}_\ell
    \rightarrow
           \frac{16 \pi^2 i S}{N (L+1)} g_{\ell + 1},
           ~~~~~{\rm for} ~~\ell \geq 3.
           \nonumber
    \end{eqnarray}
  If we perform this operation to all the vertices of all the $L$-loop diagrams with $P$ propagators, 
  we obtain the second modification to (\ref{Weffh1}).
  We denote this as $W_2^{(L)}$.
    
  Finally, we can consider various diagrams where the type I\hspace{-.1em}I\hspace{-.1em}I vertices are used.
  There are infinite number of the interaction terms in $V(\Phi)$.
  So, it is hard to compute all the diagrams, 
  but we can see that these diagram is $\mathcal{O}(1/m)$.
  Therefore, these contributions to the effective superpotential are suppressed in $\mathcal{N}=1$ limit.
  We denote their $L$-loop contributions as $W_3^{(L)}$.
  
  As a result, we have obtained the following formula \cite{IM2}:
  the contribution of the $L$-loop diagrams with $P$ propagators to the effective superpotential is
    \begin{eqnarray}
    W_{eff}^{(L)}
    &=&    N \frac{\partial F^{(L)}}{\partial S}
         + \frac{\partial^2 F^{(L)}}{\partial S^2} w^\alpha w_\alpha
           \nonumber \\
    &-&    \frac{16 \pi^2 i P \tilde{g}_3 S}{(L + 1) m \tilde{g}_2} 
           \left(
           \frac{\partial F^{(L)}}{\partial S}
           \right)
         + W_2^{(L)}
         + W_3^{(L)}.
           \nonumber
    \end{eqnarray}
  As discussed above, $F^{(L)}$ can be identified with $L$-loop contribution
  to the planar free energy of the matrix model.
  The second line of the above formula is $\mathcal{O}(1/m)$.
  Hence, we can see that, in $\mathcal{N}=1$ limit, we recover the result of \cite{DV2,DGLVZ}.
  
  We have successfully obtained the result that the effective superpotential has been deformed 
  by spontaneously broken $\mathcal{N}=2$ supersymmetry.
  However, it is hard to compute $W_3^{(L)}$ by the diagrammatical computation.
  Hence, in order to obtain the effective superpotential completely, we need other approach.
  In \cite{CDSW} and \cite{CSW1,CSW2,Ferrari1,Ferrari2}, 
  the generalized Konishi anomaly equations are used to provide the proof of \cite{DV2}.
  In our model, these equations can be used to obtain the effective superpotential
  and we can compute $W_3^{(L)}$ explicitly \cite{IM3}.

%%%%%%%%%%%%%%%%%%%%%%%%%%%%%%%%%%%%%%%%%%%%%%%%%%%%%%%%%%%%%%%%%%%%%%%%%%%%%%%%%%%%%%%%%%%%%%%%%%%%%%%%%%%%%%%%%%%%
\section*{Acknowledgements}
  We thank Hiroshi Itoyama for useful discussions.
  Support from the 21 century COE program ``Constitution of wide-angle mathematical basis focused on knots'' 
  is gratefully appreciated.

%%%%%%%%%%%%%%%%%%%%%%%%%%%%%%%%%%%%%%%%%%%%%%%%%%%%%%%%%


\end{document}